\documentclass[runningheads]{llncs}

\usepackage{algorithm}
\usepackage{algorithmic}
\usepackage{amssymb}
\usepackage{xspace}
\usepackage{latexsym}
\usepackage{url}
\usepackage{wrapfig}
\usepackage[pdftex]{graphicx}

\renewcommand{\algorithmiccomment}[1]{//#1 }

\newcommand{\dm}{Data Mining}
\newcommand{\ras}{association rule}
\newcommand{\rass}{association rules}

\newcommand{\igart}{{\sf ${\cal GART}$}}
\newcommand{\gar}{{\it RulEE-GAR}}

\begin{document}

\pagestyle{empty}

\mainmatter

\title{Using Taxonomies to Facilitate the Analysis of the Association Rules}


\author{Marcos Aur\'{e}lio Domingues\inst{1}
\and Solange Oliveira Rezende\inst{2}}


\institute{LIACC-NIAAD -- Universidade do Porto \\
Rua de Ceuta, 118, Andar 6 -- 4050-190 Porto, Portugal \\
\email{marcos@liacc.up.pt} \and 
Instituto de Ci\^{e}ncias Matem\'{a}ticas e de Computa\c{c}\~{a}o -- Universidade de S\~{a}o Paulo \\
Av. Trabalhador S\~{a}o-Carlense, 400, Cx. Postal 668 -- 13560-970 S\~{a}o Carlos, SP, Brazil \\
\email{solange@icmc.usp.br}}

\maketitle

\begin{abstract}
The Data Mining process enables the end users to analyse,
understand and use the extracted knowledge in an intelligent
system or to support in the decision-making processes. However,
many algorithms used in the process encounter large quantities of
patterns, complicating the analysis of the patterns. This fact
occurs with association rules, a Data Mining technique that tries
to identify intrinsic patterns in large data sets. A method that
can help the analysis of the association rules is the use of
taxonomies in the step of  post-processing knowledge. In this
paper, the \igart \ algorithm is proposed, which uses taxonomies
to generalize association rules, and the \gar \ computational
module, that enables the analysis of the generalized rules.
\end{abstract}

\section{Introduction}

The development of the data storing technologies has increased the
data storage capacity of companies. Nowadays the companies have
technology to store detailed information about each performed
transaction, generating large databases. This stored information
may help the companies to improve themselves and because of this
the companies have sponsored researches and the development of
tools to analyse the databases and generate useful information.

During years, manual methods had been used to convert data in
knowledge. However, the use of these methods has become expensive,
time consuming, subjective and non-viable when applied at large
databases.

The problems with the manual methods stimulated the development of
processes of automatic analysis, like the process of Knowledge
Discovery in Databases or \dm. This process is defined as a
 process of identifying valid, novel, potentially useful,
and ultimately understandable patterns in data~\cite{FAYYAD:96f}.

In the \dm \ process, the use of the association rules technique
may generate large quantities of patterns. This technique has
caught the attention of  companies and research centers
\cite{BAESENS:00}. Several researches have been developed with
this technique and the results are used by the companies to
improve their businesses (insurance policy, health policy,
geo-processing, molecular
biology)~\cite{LIU:00,KOPERSKI:00,SEMENOVA:01}.

A way to solve the problem of the large quantities of patterns
extracted by the \rass \ technique is the use of taxonomies in the
step of  post-processing
knowledge~\cite{ADAMO:01,LIU:00,SRIKANT:97}. The taxonomies may be
used to prune uninteresting and/or redundant rules
(patterns)~\cite{ADAMO:01}.

In this paper the \igart \ algorithm and the \gar \ computational
module is proposed. The \igart \ algorithm ({\it Generalization of
Association Rules using Taxonomies}) uses taxonomies to generalize
\rass. The \gar \ computational module uses the \igart \
algorithm, to generalize \rass, and provides several means to
analyze the generalized rules.

This paper is organized as following: first by presenting the
\rass \ technique and some general features about the use of
taxonomies, second by describing the \igart \ algorithm and the
\gar \ computational module. Finally the results of some
experiments performed with the \igart \ algorithm along with our
conclusion are presented.

\section{Association Rules and Taxonomies}\label{sec:ra}
An \ras \ $LHS \Rightarrow RHS$ represents a relationship between
the sets of items $LHS$ and $RHS$~\cite{AGRAWAL:94}. Each item $I$
is an atom representing the presence of a particular object. The
relation is characterized by two measures: support and confidence.
The support of a rule $R$ within a dataset $D$, where $D$ itself
is a collection of sets of items (or itemsets), is the number of
transaction in $D$ that contain all the elements in $LHS \cup
RHS$. The confidence of the rule is the proportion of transactions
that contain $LHS \cup RHS$ with respect to the number of
transactions that contain $LHS$. The problem of mining \rass \ is
to generate all \rass \ that have support and confidence greater
than the minimum support and minimum confidence defined by the
user to mine \rass. High values of minimum support and minimum
confidence just generate trivial rules. Low values of minimum
support and minimum confidence generate large quantities of rules
(patterns), complicating the user's analysis.

\begin{figure}[H]
  \centering
  \includegraphics[scale=.6]{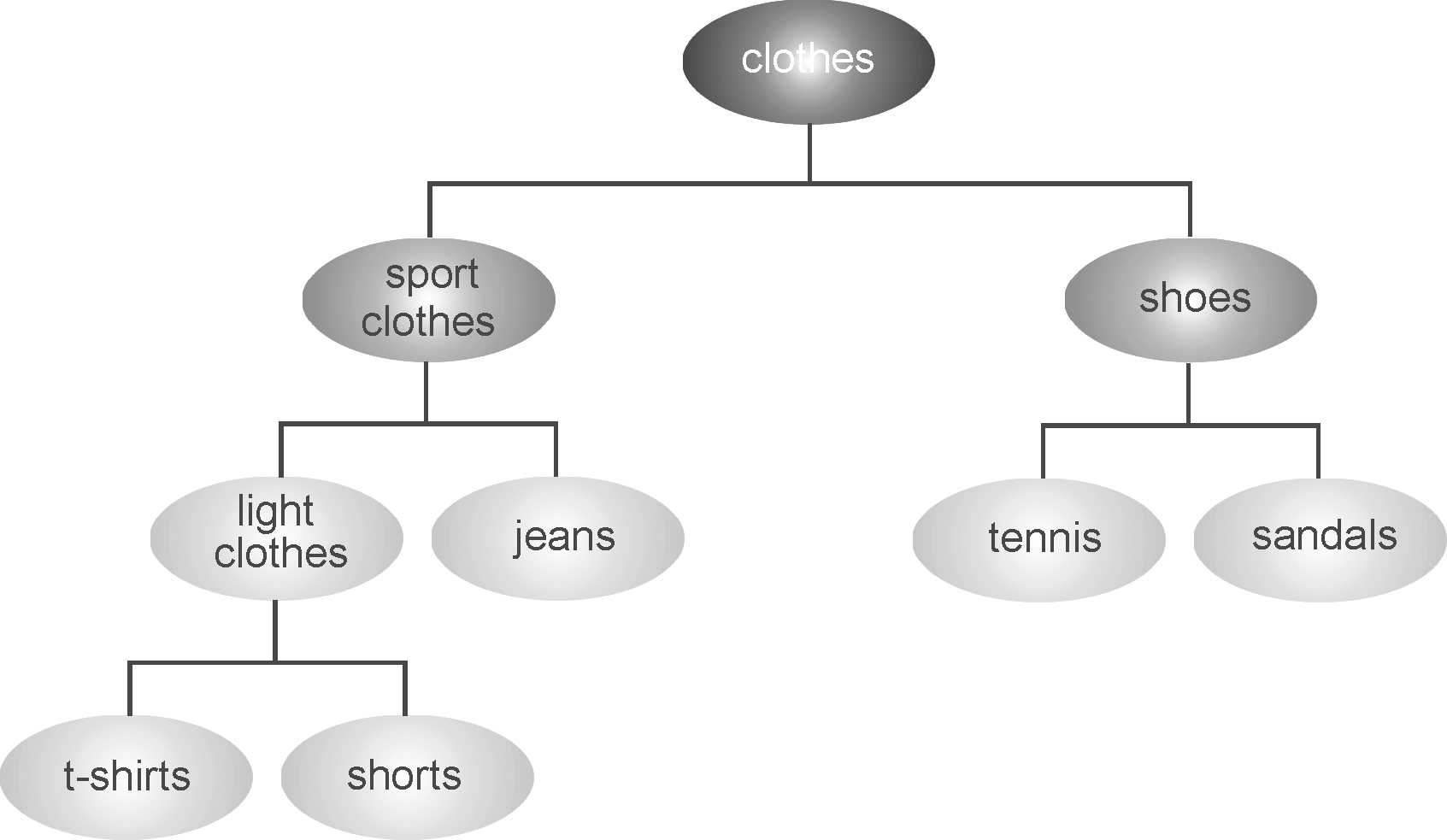}
  \caption{An example of taxonomy for clothes.}
  \label{TAXONOMIA}
\end{figure}

A way of overcoming the difficulties in the analysis of large
quantities of \rass \ is the use of taxonomies in the step of
 post-processing knowledge. The use of taxonomies may help the user
to identify interesting and useful knowledge in the extracted
rules set. The taxonomies represent a collective or individual
characterization of how the items can be classified
hierarchically~\cite{ADAMO:01}. In Fig.~\ref{TAXONOMIA} an example
of a taxonomy is presented where it can be observed that: {\it
t-shirts} are {\it light clothes}, {\it shorts} are {\it light
clothes}, {\it light clothes} are a kind of {\it sport clothes},
{\it sandals} are a kind of {\it shoes}.

In the literature there are several algorithms to generate \rass \
using taxonomies (generalized \rass). Algorithms like {\it
Cumulate} and {\it Stratify}~\cite{SRIKANT:97} generate rules sets
larger than rules sets generated without taxonomies (because they
generate \rass \ with and without taxonomies). To try decrease the
quantity of generated rules, a subjective measure is used to prune
the uninteresting rules~\cite{SRIKANT:97}. The subjective measure
does not guarantee that the quantity of rules will decrease. Our
method proposes an algorithm and a module of
post-processing~\cite{DOMINGUES:04a}. Using the module, the user
looks to a small set of rules without taxonomies, builds some
taxonomies and then uses the algorithm to generalize the \rass,
pruning the original rules that are generalized. Thus our
algorithm always decreases or keeps the volume of the rules sets.
The proposed algorithm and module are presented in
Section~\ref{al} and~\ref{mc}.

\section{The Algorithm \igart}\label{al}
\begin{figure}[H]
  \centering
  \includegraphics[scale=.5]{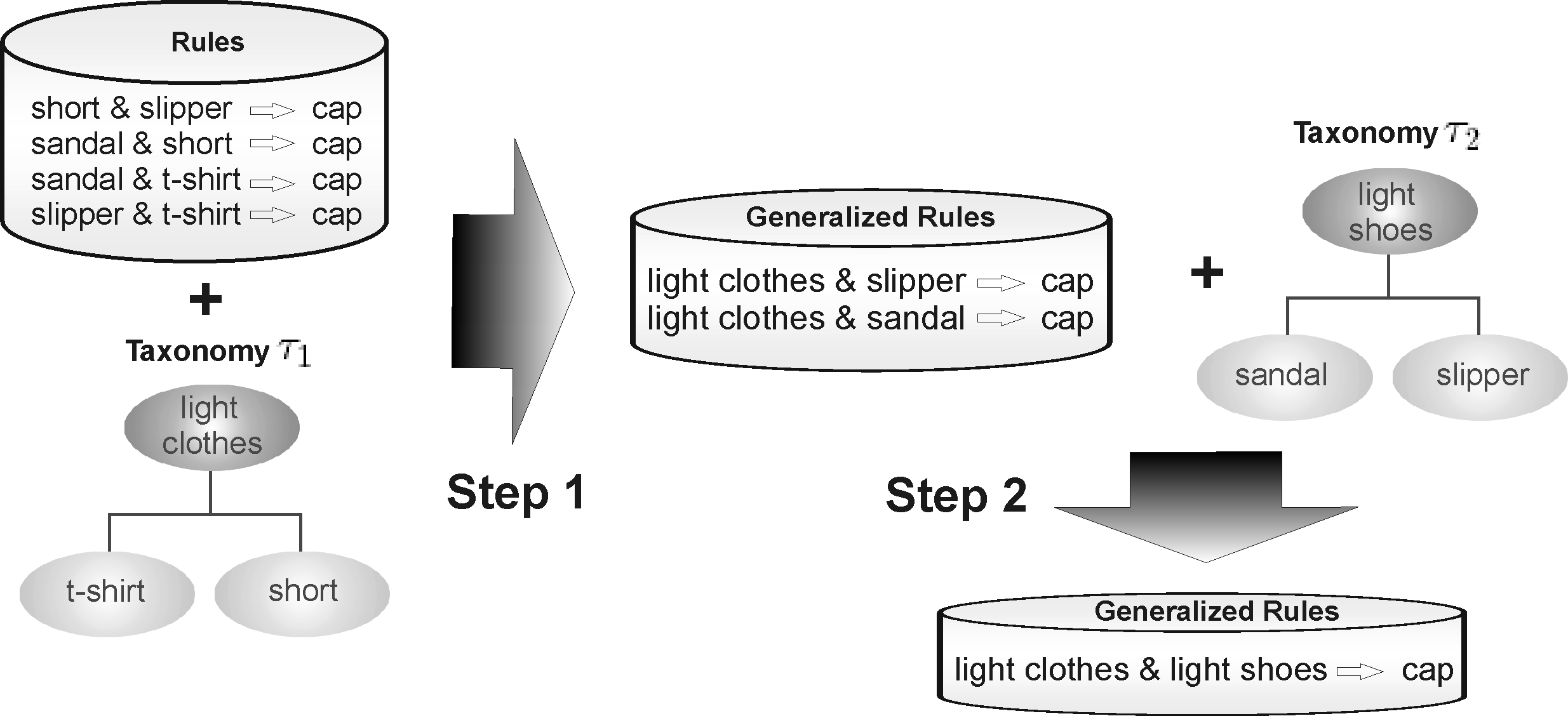}
  \caption{Generalization of \rass \ using two taxonomies.}
  \label{fig:gen2}
\end{figure}

We analysed the structure of the \rass \ generated by algorithms
that do not use taxonomies. The results of the analysis show us
that it is possible to generalize \rass \ using taxonomies. In
Fig.~\ref{fig:gen2} we show how the \rass \ can be generalized.

First we changed the items {\it t-shirt} and {\it short} of the
rules {\it short \& slipper $\Rightarrow$ cap}, {\it sandal \&
short $\Rightarrow$ cap}, {\it sandal \& t-shirt $\Rightarrow$
cap} and {\it slipper \& t-shirt $\Rightarrow$ cap} by the item
{\it light clothes} (which represents a generalization). This
change generated two rules \mbox{{\it light clothes \& slipper
$\Rightarrow$ cap}} and two rules \mbox{{\it light clothes \&
sandal $\Rightarrow$ cap}}. Next, we pruned the repeated
generalized rules, maintaining only the two rules: {\it light
clothes \& slipper $\Rightarrow$ cap} and {\it light clothes \&
sandal $\Rightarrow$ cap}.

The two rules generated by the Step 1 (Fig.~\ref{fig:gen2}) were
generalized again. We changed the items {\it slipper} and {\it
sandal} by the item {\it light shoes} (which represented another \
generalization) \ generating \ two \ rules \ \mbox{{\it light
clothes \& light shoes $\Rightarrow$ cap}}. \\ Then we pruned the
repeated generalized rules again, maintaining only one generalized
\ras: \mbox{{\it light clothes \& light shoes $\Rightarrow$ cap}}.

\begin{figure}[H]
  \centering
  \includegraphics[scale=.4]{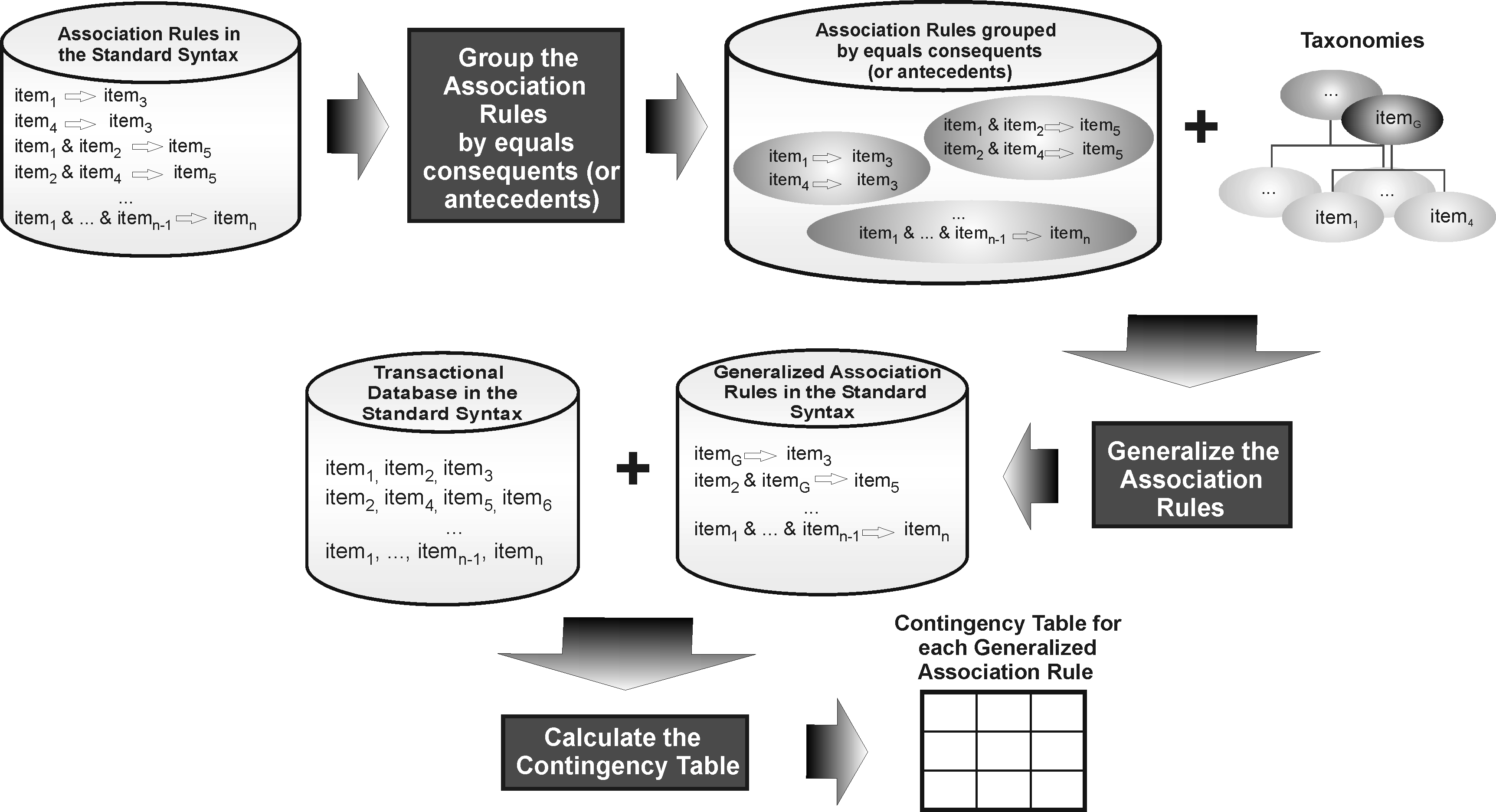}
  \caption{The proposed algorithm to generalize \rass.}
  \label{fig:proc_gen}
\end{figure}

Due to the possibility of generalization of the \rass \
(Fig.~\ref{fig:gen2}), we propose an algorithm to generalize
\rass. The proposed algorithm is illustrated in
Fig.~\ref{fig:proc_gen}. We called the proposed algorithm of
\igart \ ({\it Generalization of Association Rules using
Taxonomies}).

The proposed algorithm just generalizes one side of the \rass \ -
$LHS$ or $RHS$ (after to look to a small set of rules without
taxonomies, the user decides which side will be generalized).
First, we grouped the rules in subsets that present equal
antecedents or consequents. If the algorithm were used to
generalize the left hand side of the rules ($LHS$), the subsets
would be generated using the equals consequents ($RHS$). If the
algorithm were used to generalize the right hand side of the rules
($RHS$), the subsets would be generated using the equal
antecedents ($LHS$). Next, we used the taxonomies to generalize
each subset (as illustrated in Fig.~\ref{fig:gen2}). In the final
algorithm we stored the rules in a set of generalized \rass.

In the final algorithm, we also calculated the Contingency Table
for each generalized \rass \ to get more information about the
rules. The Contingency Table of a rule represents the coverage of
the rule with respect to the database used in its
mining~\cite{LAVRAC:99}. With the calculation of the Contingency
Table we finished the algorithm.

\section{The Computational Module \gar}\label{mc}
\begin{figure}[H]
  \centering
  \includegraphics[scale=.41]{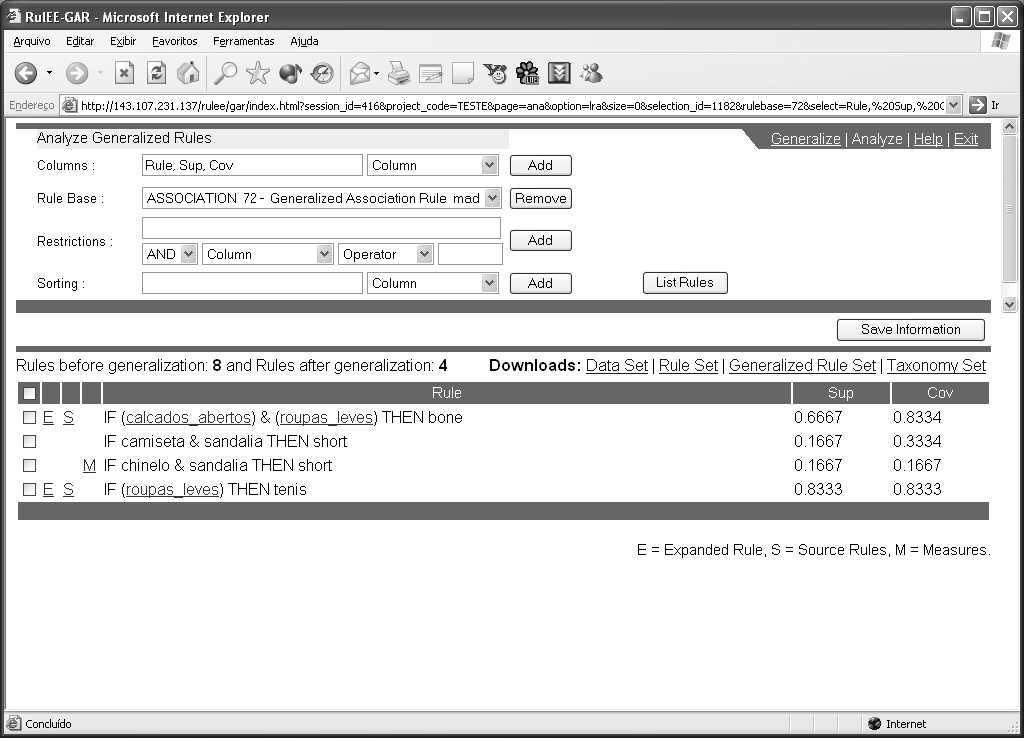}
  \caption{Screen of the analysis interface of generalized \rass.}
  \label{fig:gar_ana1}
\end{figure}

In this section we present the \gar \ computational module that
provides means to generalize \rass \ and also to analyze the
generalized rules~\cite{DOMINGUES:04a}. The generalization of the
\rass \ is performed by the \igart \ algorithm, described in the
previous section. Next we describe the means to analyze the
generalized \rass. In Fig.~\ref{fig:gar_ana1} we showed the screen
of the interface that enables the user to analyze and to explore
the generalized rules sets.

On the screen of the analysis interface of generalized rules
(Fig.~\ref{fig:gar_ana1}) there are some spaces where the user
puts data to make a query and select a set of generalized rules,
accompanied or not of several evaluation
measures~\cite{LAVRAC:99}, to be analyzed. Besides allowing the
user to select a set of rules, the interface provides four links
in the section {\it Downloads} to look for  and/or  download  the
files. The files contain, respectively, the set of transactional
data ({\it Data Set}), the set of source rules ({\it Rule Set}),
the set of generalized rules ({\it Generalized Rule Set}) and the
set of taxonomies used to generalize the rules ({\it Taxonomy
Set}).

Besides links for visualization and/or download of the files, each
generalized \ras \ presents others links that enable the user to
explorer information about the generalization of the rule. The
links are positioned at the left side of the rules
(Fig.~\ref{fig:gar_ana1}). The links are described as following:

\begin{description}
\item[\textit{Expanded Rule}] It is represented in the interface
by the letter ``E''. This link enables the user to see the
generalized rule in expanded way. The generalized items of a rule
are changed by the respective specific items.

\item[\textit{Source Rules}] It is represented in the interface by
the letter ``S''. This link enables the user to see the source
rules that were generalized.

\item[\textit{Measures}] It is represented in the interface by the
letter ``M''. This link is available only if the user selects the
support ({\it Sup}) and/or confidence ({\it Cov}) measures in its
query and these measures present values lower than the minimum
support and/or minimum confidence values defined to the mining
process of the rules set not generalized. With this link it is
possible to see which generalized rules have support and/or
confidence values lower than the minimum support and/or minimum
confidence values.
\end{description}

In Fig.~\ref{fig:gar_ana1} we also see that the generalized items
in a rule (items between parentheses) are presented as links.
These links enable the user to see the source items that were
generalized. In the analysis interface, the user can also store
the information, selected by the query, in a text file.

\section{Experiments}\label{ex}
We performed some experiments using the \igart \ algorithm to
demonstrate that the use of taxonomies, to generalize large rules
sets, reduces large quantities of \rass \ and makes easy the
analysis of the rules.

The experiments were performed using a sale database of a
Brazilian supermarket. The database contained sales data of the
recent 3 month. We made 4 partitions of the database to perform
the experiments. The partitions were made using the sale data
along of 1 day, 7 days, 14 days and 1 month.

To generate the \rass, we used the implementation of the {\it
Apriori} algorithm performed by Chistian
Borgelt\footnote{Available for downloading at the web site
\url{http://fuzzy.cs.uni-magdeburg.de/~borgelt/software.html}.}
with minimum support value equal 0.5, minimum confidence value
equal 0.5 and a maximum number of 5 items by rule. The generated
rules sets are described as following:

\begin{itemize}
\item RuleSet\_1day - 32668 rules generated using the partition of
1 day; \item RuleSet\_7days - 19166 rules generated using the
partition of 7 days; \item RuleSet\_14days - 16053 rules generated
using the partition of 14 days; \item RuleSet\_1month - 21505
rules generated using the partition of 1 month; \item
RuleSet\_3months - 19936 rules generated using the whole database
(3 months of sale data).
\end{itemize}

\begin{figure}[H]
  \centering
  \includegraphics[scale=.65]{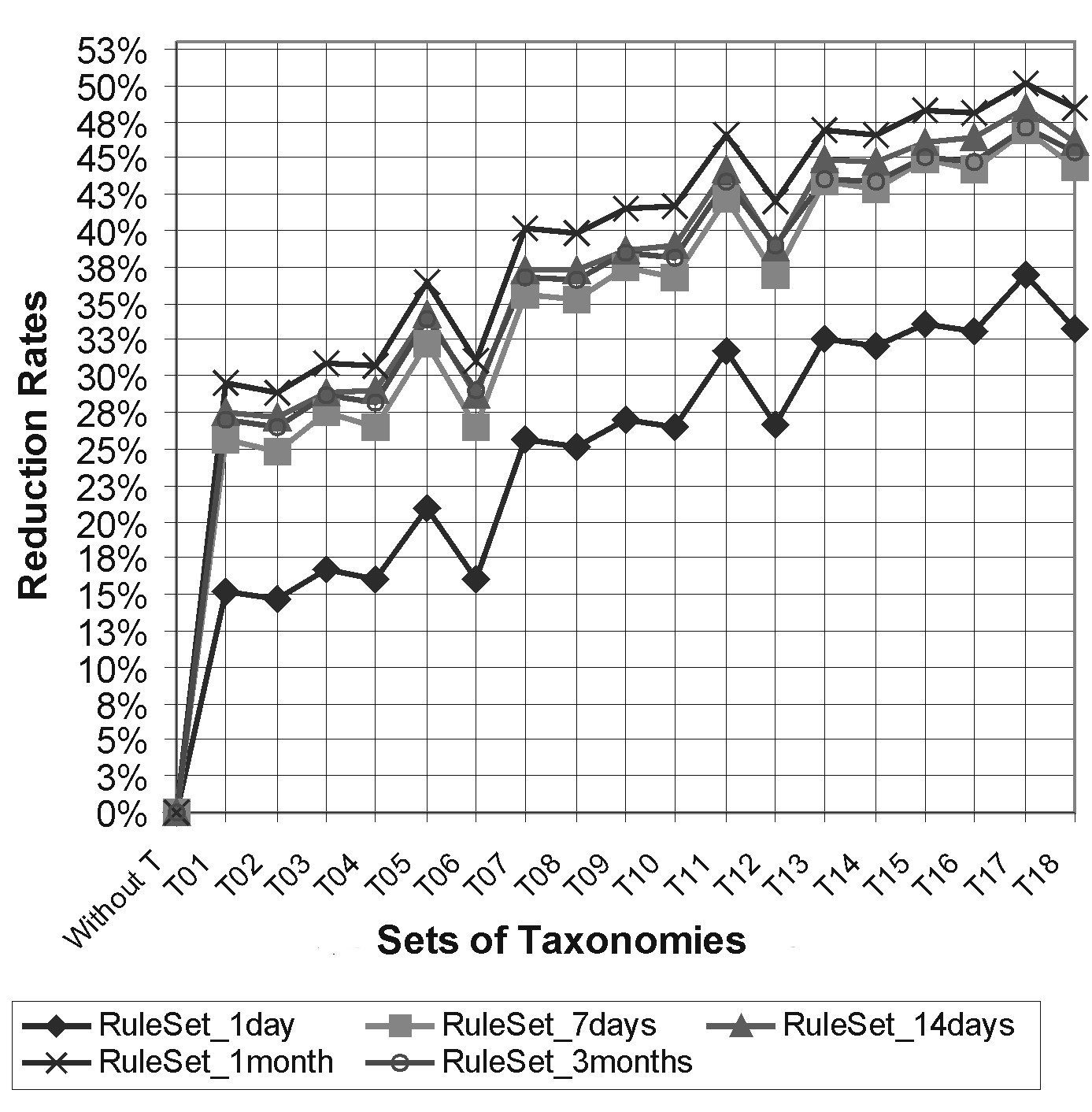}
  \caption{Reduction rates got using taxonomies to generalize \rass.}
  \label{fig:grafico1}
\end{figure}

To perform the experiments, we looked to the database and to the 5
sets of \rass \ generated to make 18 sets of taxonomies. Then we
ran the \igart \ algorithm combining each set of taxonomies with
each set of rules. In Fig.~\ref{fig:grafico1} a chart is presented
that shows the reduction rates of the 5 rules sets after running
\igart \ algorithm using the 18 sets of taxonomies to generalize
each rules set. In Fig.~\ref{fig:grafico1} the sets of taxonomies
are called ``T'' followed by an identification number, as for
example: T01.

As  it can be observed in Fig.~\ref{fig:grafico1}, the experiments
show reduction rates of the sets of \rass \ varying from 14,61\%
to 50,11\%.

\section{Conclusion}\label{cf}
A problem found in the \dm \ process is the fact that several of
the used algorithms generate large quantities of patterns,
complicating the analysis of the patterns. This problem occurs
with the \rass, a \dm \ technique that tries to identify all the
patterns in a database.

The use of taxonomies, in the step of knowledge post-processing,
to generalize and to prune uninteresting and/or redundant rules
may help the user to analyze the generated \rass.

In this paper we proposed the \igart \ algorithm that uses
taxonomies to generalize \rass. We also proposed the \gar \
computational module that uses the \igart \ algorithm to
generalize \rass \ and provides several means to analyse the
generalized \rass. Then we presented the results of some
experiments performed to demonstrate that the \igart \ algorithm
may reduce the volume of the sets of \rass. As the sets of
taxonomies were made by the user, others sets of taxonomies may
generate reduction rates higher than the rates presented in our
experiments, mainly whether the sets were made by
experts in the application domain.\\

\noindent {\bf Acknowledgements.} This work was supported by the
Funda\c{c}\~{a}o de Amparo \`{a} Pesquisa do Estado de S\~{a}o
Paulo (FAPESP), Brazil.
%
%
\bibliographystyle{plain}
\bibliography{root}

\end{document}